\documentclass[useAMS,usenatbib]{mn2e}
\usepackage{graphicx}

\newcommand{\kms}{~km~s$^{-1}$}
\newcommand{\apj}{ApJ}

\def\cm3{~cm$^{-3}$ }
\newcommand{\dotM}{~M$_{\odot}$~yr$^{-1}$~}

\newcommand{\wbb}{NGC~2359~}
\newcommand{\wbbe}{NGC~2359}
\newcommand{\XMM}{{\it XMM-Newton~}}

\title[]{X-rays from wind-blown bubbles: an \XMM detection of \wbb}
\author[S.A.Zhekov]{Svetozar A. Zhekov
\thanks{E-mail: szhekov@space.bas.bg} \\
Space Research and Technology Institute, Akad. G.
Bonchev str., bl.1, Sofia 1113, Bulgaria\\
}

\date{}

\pagerange{\pageref{firstpage}--\pageref{lastpage}} \pubyear{2002}

\begin{document}

\maketitle

\label{firstpage}

\begin{abstract}
We present an analysis of the \XMM observation of the wind-blown
bubble \wbbe. This is the first detection of this object in X-rays.
The X-ray emission of \wbb is soft and originates from a thermal 
plasma with a typical temperature of kT~$\sim 0.2$~keV. 
A direct comparison between the one-dimensional hydrodynamic model of 
wind-blown bubbles and the X-ray spectrum of \wbb suggests a reduced 
mass-loss rate of the central star in order to provide
the correct value of the observed flux.
The central star of the nebula, WR 7, is an X-ray source. Its 
emission is similar to that of other presumably single Wolf-Rayet
stars detected in X-rays. The WR 7 spectrum is well represented
by the emission from a two-temperature plasma with a cool component of
kT~$\sim 0.6$~keV and a hot component of kT~$\sim 2.7$~keV. 

\end{abstract}


\section{Introduction}

The currently accepted physical picture for the origin of the
optical nebulosities around massive stars of early spectral type (O,
Of, and Wolf-Rayet (WR)) assumes that they are formed from
the interaction of the stellar wind with circumstellar gas. The
flow pattern, resulting from such an interaction, was first recognized
by \citet{pi_68}. These so called wind-blown bubbles (WBB) consist of
two regions of shocked gas: the outer one is filled with the shocked
circumstellar gas and the inner one (the hot bubble) harbours the 
shocked stellar wind. A contact discontinuity separates these regions 
and the plasma in the hot bubble can cool down due to the high
efficiency of electron thermal conduction 
(e.g., \citealt{weaver_77}).

The physics of the hot interior of a WBB
is a cornerstone for the physical picture of these objects. The hot
bubble is the place where the stellar wind energy is supplied and then
used for driving the entire structure. Because of the high velocity
(1000 - 3000\kms) of the stellar winds of massive O and WR stars,
and thus the high shock velocities in WBBs, the expected plasma
temperatures are very high. Therefore, WBBs are expected to
emit X-rays and the key characteristic of their X-ray emission is
that it should be spatially located {\it inside} the optical nebula.

\citet{boch_88} reported the first successful detection of X-rays 
from a WBB: the observation of NGC~6888 by {\it Einstein} revealed 
hot plasma with a typical temperature of kT~$= 0.28 - 0.8$~keV 
(90\% confidence interval). Subsequent studies of this object with 
{\it ROSAT} \citep{wri_94}, {\it ASCA} \citep{wri_05}, 
{\it Chandra} (\citealt{chu_06}; \citealt{toala_14}), and {\it Suzaku}
\citep{zhp_11} provided further details about the X-ray
emission from NGC~6888 and established that the plasma temperature of
the hot bubble is kT~$=0.12-0.15$~keV. The {\it Suzaku} observations of
NGC~6888 provided us with the first X-ray spectrum with good photon
statistics of an entire WBB. Its analysis gave the first observational
evidence about the origin of the hot gas: the WBB interior is
filled mostly by gas that flowed into the hot bubble from the optical
nebula (Zhekov \& Park 2011). This, along with the relatively low
plasma temperature of the hot bubble, is a solid indication for
efficient electron thermal conduction operating in WBBs.

Unfortunately, there are not many detections of X-ray emission from
WBBs. There is only one more object, S308, that has been detected 
and studied in some detail. Analysis of the {\it ROSAT} \citep{wri_94}
and the \XMM \citep{chu_03} observations of S308 showed
that its X-ray spectrum is soft with plasma temperature of kT~$=
0.15$~keV. Recent analysis of new \XMM observations that completed 
coverage of this WBB observed in X-rays to 90\% of its spatial extent 
confirmed the basic properties of the hot plasma in S308
\citep{toala_12}. It also provided evidence that the chemical 
composition of the hot gas is similar to that of the optical nebula 
which, as in the case of NGC~6888, indicates that electron thermal 
conduction plays an important role for the physics of this object 
as well.

Apart from the two X-ray detections just mentioned, there are two
reports in the literature of non-detections of WBBs in X-rays, namely,
of the nebulae around the WR stars WR 16 \citep{to_gu_13}
and WR 40 \citep{goss_05}. Thus, any new data on the X-ray
emission from WBBs are of great importance in helping us to better
understand the physics of these objects.

\begin{figure*}
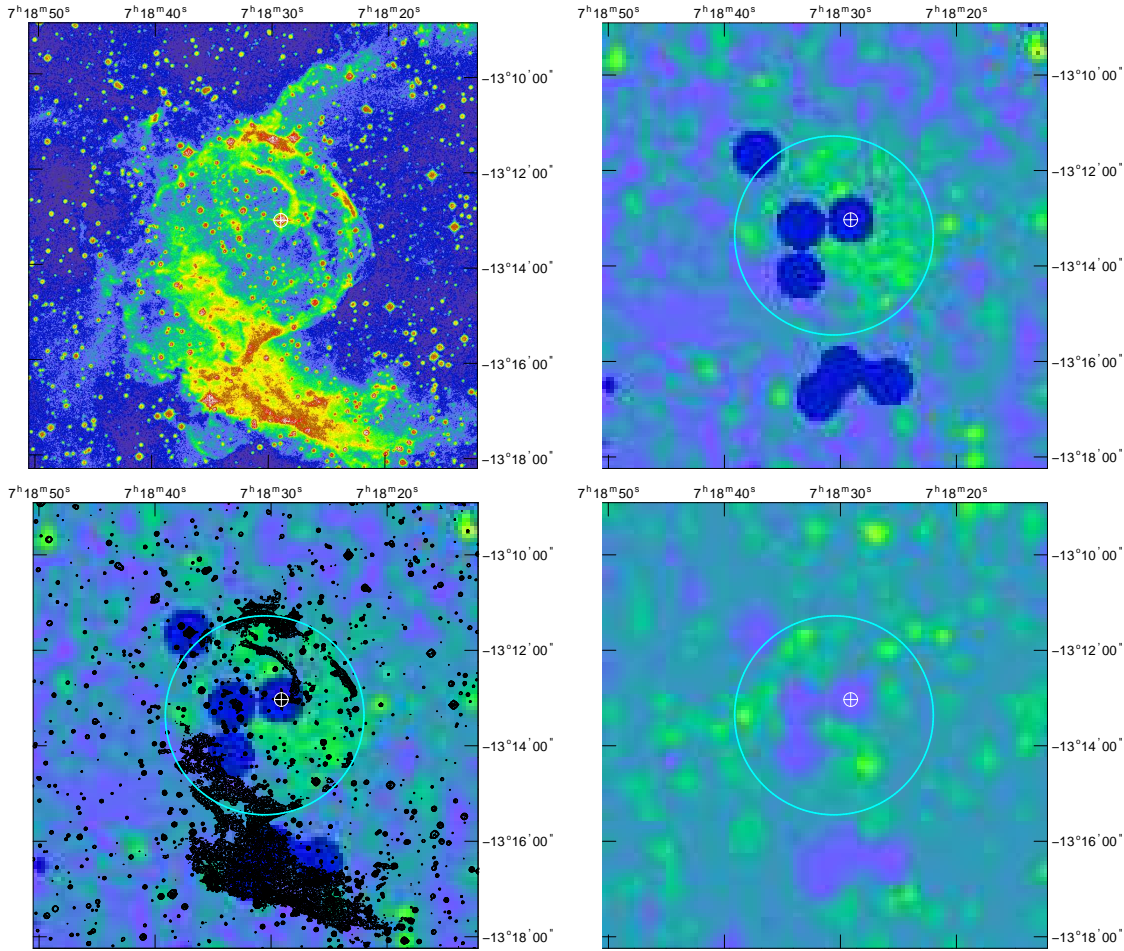

 \centering\includegraphics[width=2.94in,height=2.5in]{fig1a.eps}
 \centering\includegraphics[width=2.94in,height=2.5in]{fig1b.eps}
 \centering\includegraphics[width=2.94in,height=2.5in]{fig1c.eps}
 \centering\includegraphics[width=2.94in,height=2.5in]{fig1d.eps}
\caption{
Images of \wbb in
the optical (upper left panel),
the soft (0.4 - 1.25 keV) X-rays (upper right panel),
the soft X-rays overlaid with the contours of the optical image
(lower left panel),
the hard (2.0 - 4.0 keV) X-rays (lower right panel).
The position of the central star of the WBB is marked by a crosshairs
symbol. The circle in light blue colour ($\sim 2.1$ arcmin in radius)
is drawn to facilitate the comparison between different images.
}
\label{fig:images}
\end{figure*}

In this paper, we report results from the \XMM observation of the WBB 
\wbbe, the third object of this kind detected in
X-rays. The paper is organized as follows. We give some basic
information about \wbb in Section~\ref{sec:ngc2359}. In
Section~\ref{sec:data}, we review the \XMM observation. In
Section~\ref{sec:xray}, we present results from analysis of the X-ray
properties of \wbb and the two brightest point sources within the 
optical nebula. In Section~\ref{sec:discussion}, we discuss our
results and we present our conclusions in
Section~\ref{sec:conclusions}.

\section{The wind-blown bubble \wbb}
\label{sec:ngc2359}

The nebula \wbb around the WR star WR 7 is a classical wind-blown
bubble with angular size of 4.5 arcmin \citep{chu_83}. The
expansion velocity of the nebula is not well constrained and a typical
range for its value is 10-30\kms (see the discussion in
\citealt{chu_88}).
The distance to \wbb(WR 7) is d~$= 3.67$~kpc \citep{vdh_01},
thus, the dynamical age (radius divided by the expansion velocity)
of the nebula is $\sim 78,500 - 236,000$~years.
No X-ray detection of \wbb has been previously reported.

The central star, WR 7, is a WN4 star and the optical extinction 
toward it is A$_{\mbox{V}} = 2.14$~mag (\citealt{vdh_01}; 
A$_v = 1.11$ A$_{\mbox{V}}$)
implying a foreground column density of
N$_H = (3.42-4.75)\times10^{21}$~cm$^{-2}$.
The range corresponds to the conversion that is used:
N$_H = (1.6-1.7)\times10^{21}$A$_{\mbox{V}}$~cm$^{-2}$
(\citealt{vuong_03}, \citealt{getman_05});
and 
N$_H = 2.22\times10^{21}$A$_{\mbox{V}}$~cm$^{-2}$
\citep{go_75}.
We adopt the stellar wind
parameters (velocity and mass loss) of
V$_{wind} = 1600$\kms and 
$\dot{M} = 4\times10^{-5}$\dotM \citep{ham_koe_98}.

\section{Observations and data reduction}
\label{sec:data}

\wbb was observed with \XMM on 2013 Apr 9 (Observation ID 0690390101)
with a nominal exposure of $\sim 111$~ksec. Because of the anticipated
faint X-ray emission from \wbbe, our analysis is focused on the data
from the European Photon Imaging Camera (EPIC) having one pn and
two MOS detectors\footnote{see \S~3.3 in the \XMM Users Handbook,
http://xmm.\\esac.esa.int/external/xmm\_user\_support/documentation/uhb
}.
For the data reduction, we made use of the \XMM 
SAS\footnote{Science Analysis Software, http://xmm.esac.esa.int/sas}
12.0.1 data analysis software. For the analysis of the EPIC MOS and pn
observations, we also made use of the \XMM Extended  Source Analysis 
Software (XMM-ESAS) 
package\footnote{http://xmm.esac.esa.int/sas/current/doc/esas},
which is now incorporated in SAS.

Before proceeding with the spectral extraction, we verified
that the studied object was detected. For that purpose, we constructed
an X-ray image following the recipe described in the XMM-ESAS 
Cookbook\footnote{ftp://xmm.esac.esa.int/pub/xmm-esas/xmm-esas.pdf}.
First, we created filtered event files using the ESAS commands
{\it mos-filter} and {\it pn-filter} which minimizes the contamination 
of the soft proton flaring in the EPIC data in a robust manner. 
The relatively strong point sources were excised as well.
Then, we constructed quiescent background spectra and images. We used
the corresponding results and the XMM-ESAS command {\it adapt\_900} to 
construct a combined EPIC background-subtracted and exposure-corrected 
image.

Figure~\ref{fig:images} presents the background-subtracted images of
\wbb in the soft (0.4 - 1.25 keV) and hard (2.0 - 4.0 keV) energy
bands along with an optical image of the studied object. A minimum of
20 counts per bin were requested for the smoothed images. As seen
from these results, soft X-ray emission is detected from \wbbe, and
this emission originates from regions {\it inside} the optical nebula.

For the analysis of the X-ray spectrum of \wbbe, we decided to adopt
the standard approach (i.e., annular background subtraction) 
instead of following the one described in the
XMM-ESAS Cookbook for strong extended sources. The reason is
that the X-ray emission from the region of \wbb is dominated by 
the background emission (instrumental and cosmic). So, a global fit to
the total X-ray emission that consists of various background
components and the emission of \wbb will simply `wash out' the
spectral parameters of the latter. 
In addition, the \wbb X-ray emission is of limited angular extent
making an annular background spectrum reasonable.

\begin{figure}
 \centering\includegraphics[width=2.8325in,height=2.5in]{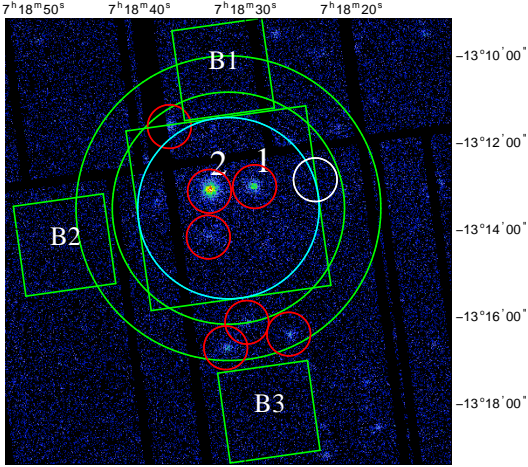}
\caption{
The raw EPIC-pn image of \wbb in the (0.2 - 10 keV) energy band with
the spectral extraction regions.
The source (\wbbe) extraction region is marked with the central square
(the circle in light blue is identical to the one shown in
Fig.~\ref{fig:images}). The basic background
extraction region is denoted by the large annulus. Additional
background regions are given with small squares 
and labelled `B1', `B2' and `B3'. The
point sources that were excised from the source and background regions
are shown with small circles (in red). The two relatively strong point
sources analysed in this study are labelled by `1' (the central star
WR 7) and `2' (the object [S3XJ067536]).
The white circle denotes the background region for sources `1'
and `2'.
}
\label{fig:regions}
\end{figure}

Figure~\ref{fig:regions} shows the
X-ray image of \wbb and the corresponding regions for the
spectral extraction. We note that our basic background region is that
denoted by the circular annulus. We do so as the X-ray spectrum 
from such a region could likely combine the spectral characteristics 
of the entire \wbb background. The X-ray spectra of the other 
background regions are chosen as representative of different parts of 
the X-ray sky and of the X-ray detectors. 
The spectral extractions were done from the filtered event files (see
above) having effective exposures of 64.5 ksec (pn), 88.6 ksec (MOS1) 
and 92.8 ksec(MOS2). We used only the pn spectrum of \wbb ($\sim 970$
source counts) since the MOS spectra have considerably lower photon
statistics due to the MOS much lower effective area compared to that 
of the pn detector (see \S 3.2.2 in the \XMM Users Handbook:
footnote$^1$).

We extracted the X-ray spectra of the two brightest sources in the
region of \wbb(Fig.~\ref{fig:regions}): 
the central star of the nebula WR 7 and the HST 
GSC\footnote{The Hubble Space Telescope Guide Star Catalog,
http://gsss.stsci.edu/webservices/GSC2/GSC2WebForm.aspx}
object S3XJ067536 (the GSC object coordinates are within 1.2 arcsec 
from the X-ray coordinates as derived from the \XMM data, thus, we 
will denote this object as [S3XJ067536]). 
Since the spectra of point
sources are not background-dominated, we extracted a pn and two MOS
spectra for each object. The MOS1 and MOS2 spectra are practically
identical which allowed us to construct a total (combined) MOS 
spectrum for each point source. The resultant spectra of WR 7 had 
$\sim 865$~cts (pn) and $\sim 740$~cts (MOS), while those for
[S3XJ067536] had $\sim 2988$~cts (pn) and $\sim 2475$~cts (MOS).

Finally, we constructed response matrix files and ancillary response 
files for each spectrum by making use of the SAS procedures {\it rmfgen} 
and {\it arfgen}, respectively. For the case of \wbbe, we followed the
approach recommended for extended sources 
(e.g., by creating an image
of the source region, see \S 4.8.5 in the SAS
Users Guide\footnote{http://xmm.esac.esa.int/external/xmm\_user\_support/\\documentation/sas\_usg/USG/}).
We made use of standard as well as custom models in version 11.3.2 of
XSPEC \citep{Arnaud96} for the spectral analysis.

\section{X-ray properties}
\label{sec:xray}

\subsection{\wbb}
\label{subsec:ngc2359}

A basic characteristic of the X-ray emission from \wbb is that it is
rather soft and there is no apparent emission at energies above 2 keV 
(see Fig.~\ref{fig:images}). We note that because of the relatively
large spatial extent of the nebula and its weak X-ray emission, it is
not possible to correct the background-subtracted spectrum for the
instrumental Al K$_\alpha$ line very well. For these reasons, our
spectral analysis is focused in the 
(0.3 - 1.44 keV)
energy range.
To improve the photon statistics, we
re-binned the \wbb spectrum to have a minimum of 50 counts per bin.

We used the discrete-temperature plasma model $vapec$ in XSPEC 
to derive the global properties of the X-ray emission from \wbbe: 
typical plasma temperature, X-ray absorption, observed flux etc. 
Since the quality of the spectrum is not very high, our model fits
assumed fixed values for the elemental abundances of the X-ray emitting
plasma. As a basic set of abundances, we adopted those typical for the
optical nebula of \wbb (see Table 3 in \citealt{este_93}): He~$= 1$, 
C~$= 0.358$, N~$= 0.179$, O~$= 0.188$, Ne~$= 0.398$, S~$= 0.309$.
Since no values are available for Si, Ar, Ca, Fe and Ni, we adopted
the same value as for the sulphur abundance and only the 
Mg abundance was allowed to vary to improve 
the quality of the fit near 1.3-1.4 keV. All the abundance values are 
given with respect to the solar abundances of \citet{an_89}.

\begin{table}
\caption{\wbb: Spectral Model Results}
\label{tab:ngc2359}
\begin{tabular}{ll}
\hline
Parameter & 1T vapec  \\
\hline
$\chi^2$/dof  & 33/46 \\
N$_{H}$ (10$^{21}$ cm$^{-2}$) & 4.77$^{+1.49}_{-1.32}$  \\
kT (keV) & 0.21$^{+0.04}_{-0.04}$  \\
EM ($10^{55}$~cm$^{-3}$) &  11.0 \\
Mg  & 3.32$^{+1.06}_{-0.90}$  \\
F$_X$ ($10^{-14}$ ergs cm$^{-2}$ s$^{-1}$)  &
           2.33 (33.0) \\
\hline
\end{tabular}

\vspace{0.5cm} 
Note -- Results from  the fit to \XMM spectrum of \wbbe.
Tabulated quantities are the neutral hydrogen absorption column
density (N$_{H}$), plasma temperature (kT),
emission measure ($\mbox{EM} = \int n_e n_H dV $),
the X-ray flux (F$_X$) in the 0.3 - 1.5 keV range followed in 
parentheses by the unabsorbed value.  Only the abundance of Mg  
was varied in the fit and the values of other abundances were kept
fixed to their adopted values (see text in \S~\ref{subsec:ngc2359}).
Errors are the $1\sigma$ values from the fit.
\end{table}

\begin{figure}
\centering\includegraphics[width=2.25in,height=3.01338in,angle=-90]
{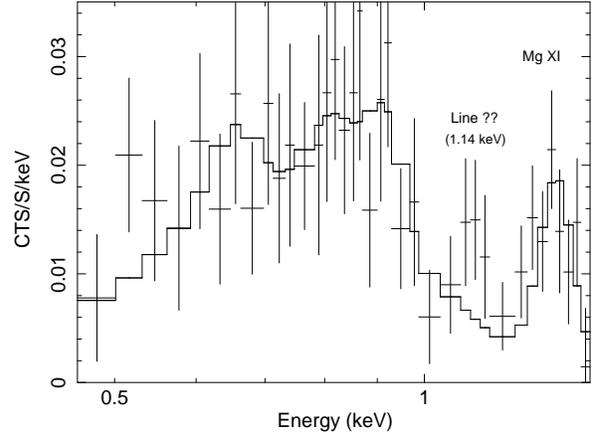}
\caption{
The \wbb background-subtracted spectrum overlaid with the
best-fit one-temperature optically-thin plasma model 
(Table~\ref{tab:ngc2359}). 
The `Mg XI' label denotes the Mg XI He-like triplet at $\sim 1.35$~keV.
For presentation purposes, the spectrum was slightly re-binned with
respect to the original binning used in the fit (see
\S~\ref{subsec:ngc2359}).
}
\label{fig:wbbspec}
\end{figure}

Figure~\ref{fig:wbbspec} and Table~\ref{tab:ngc2359} present the results
from the model fit to the X-ray spectrum of \wbbe. We note that the
X-ray emission of this object is represented well by the spectrum of
an optically-thin plasma with a relatively low temperature.
Also, the derived value of the 
X-ray absorption is consistent with the optical extinction to \wbbe:
N$_{H}(\mbox{X-ray}) = 4.77^{+1.49}_{-1.32}\times10^{21}$ vs.
N$_H (\mbox{opt.}) = (3.42-4.75)\times10^{21}$~cm$^{-2}$
(see \S~\ref{sec:ngc2359}).

Although we believe that our `basic' background spectrum (extracted
from the annulus shown in Fig.~\ref{fig:regions}) represents the
background contribution to the \wbb spectrum in the most realistic
way, we nevertheless checked our fit results by making use of other
background spectra (extracted from the square regions labelled `B1', 
`B2', `B3' in Fig.~\ref{fig:regions}). The results from these fits 
confirmed our basic findings (given in Table~\ref{tab:ngc2359}) within 
the fits uncertainties. Namely, the plasma temperature was in the 
range kT~$= (0.23 - 0.31)$~ keV and the X-ray absorption was 
N$_H = (2.4 - 6.4)\times10^{21}$~cm$^{-2}$. 

We are therefore confident in our conclusion that a relatively `cool'
plasma dominates the X-ray emission from \wbb as is the cases of S308
(e.g., \citealt{chu_03}; \citealt{toala_12}) and NGC~6888 (e.g.,
\citealt{zhp_11}).

Finally, 
we mention that the strength of the suspected line feature 
at 1.14 keV (see Fig.~\ref{fig:wbbspec}) depends on the background 
spectrum at hand, and in some cases it can even `disappear'. To
compare, the Mg XI line is always present. We could not match this 
line if the Ne and Fe abundances varied in the spectral fits although 
there are a lot of spectral lines of Ne and Fe near 1.14 keV. We thus 
believe that this line is not an intrinsic feature of the \wbb 
spectrum.

\subsection{Bright point sources}
\label{subsec:ps}

We analysed the spectra of the two relatively bright point sources in
the field of \wbbe. We re-binned the spectra to have a minimum of 20 
and 30 counts per bin for the WR 7 and [S3XJ067536] spectra, 
respectively. For each object, we fitted the pn and total MOS spectra
simultaneously.

\begin{table}
\caption{Point Sources: Spectral Model Results}
\label{tab:ps}
\begin{tabular}{lll}
\hline
Parameter & WR 7  & [S3XJ067536] \\
\hline
$\chi^2$/dof  & 87/94 & 111/180 \\
N$_{H}$ (10$^{21}$ cm$^{-2}$) & 3.44$^{+0.54}_{-0.50}$ &
                                5.89$^{+0.41}_{-0.33}$  \\
kT$_1$ (keV) & 0.60$^{+0.05}_{-0.03}$ & ... \\
kT$_2$ (keV) & 2.68$^{+1.43}_{-0.72}$ & ... \\
EM$_1$ ($10^{53}$~cm$^{-3}$) &  9.97  & ... \\
EM$_2$ ($10^{53}$~cm$^{-3}$) &  3.25  & ... \\
$\Gamma$  & ... & 1.66$^{+0.05}_{-0.04}$   \\
norm$_{pow} (10^{-5}$) &... & 5.61 \\
F$_X$ ($10^{-14}$ ergs cm$^{-2}$ s$^{-1}$)  &
           2.68 (7.27) & 23.3 (36.0)\\
F$_{X,1}$ ($10^{-14}$ ergs cm$^{-2}$ s$^{-1}$)  &
           1.90 (6.07)  & ... \\
\hline
\end{tabular}

\vspace{0.5cm}
Note -- Results from  the fits to the \XMM  spectra of the point 
sources WR 7 (a two-temperature optically-thin plasma model) and 
[S3XJ067536] 
(a power-law model). Tabulated quantities are the neutral hydrogen 
absorption column density (N$_{H}$), plasma temperature (kT$_{1,2}$),
emission measure ($\mbox{EM}_{1,2} = \int n_e n_{He} dV $),
photon power-law index ($\Gamma$; F$_X \propto E^{-(\Gamma-1)}$), 
normalization for the power-law
model (norm$_{pow}$ in units of cts keV$^{-1}$ cm$^{-2}$ s$^{-1}$ at
1 keV),
the X-ray flux (F$_X$) in the 0.3 - 8 keV range followed in
parentheses by the unabsorbed value and the flux for the cool
component for the 2T plasma model (F$_{X,1}$). Only the abundances 
of N and Mg were varied in the 2T model fit 
(N~$= 2.68^{+1.49}_{-1.86}$, Mg~$= 1.43^{+0.36}_{-0.27}$)
and the values of other abundances were kept fixed to their adopted 
values (see text in \S~\ref{subsec:ps}).
Errors are the $1\sigma$ values from the fits.
\end{table}

\begin{figure*}
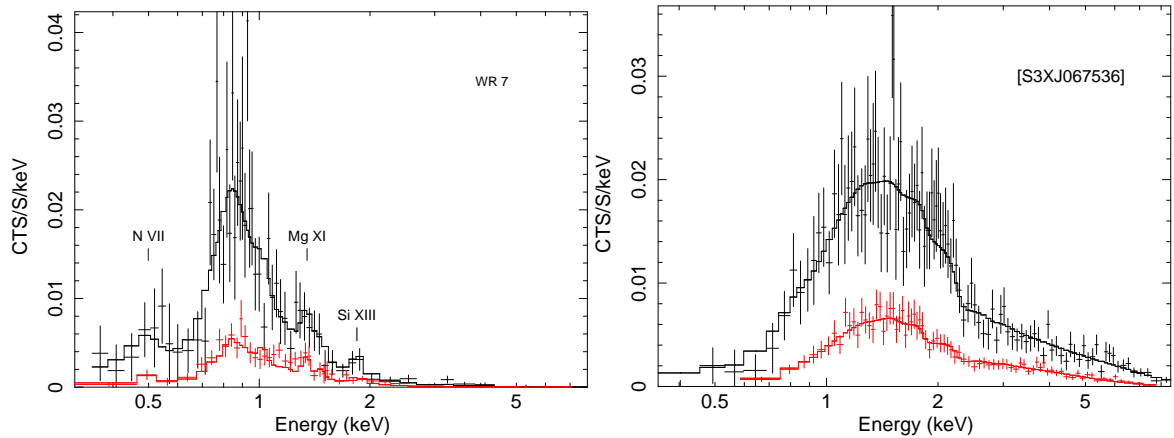

\centering\includegraphics[width=2.25in,height=3.01338in,angle=-90]
{fig4a.eps}
\centering\includegraphics[width=2.25in,height=3.01338in,angle=-90]
{fig4b.eps}
\caption{
The background-subtracted spectra of WR 7  and [S3XJ067536] overlaid
with the best-fit models (Table~\ref{tab:ps}). The pn spectrum is in
black and the MOS spectrum is in red (lower curve in each panel). Some
line features are denoted in the WR 7 spectra.
}
\label{fig:ps}
\end{figure*}

{\it WR 7.}
As in the previous studies of the X-ray emission from presumably 
single WN stars (\citealt{sk_10}, \citealt{sk_12}), we used 
an absorbed two-temperature optically-thin plasma model ({\it vapec}) 
in XSPEC to model the spectra of this WN4 star. Similarly, we adopted 
the canonical WN abundances \citep{vdh_86}. In the fit,
all abundances were kept fixed to their generic values and only the N
and Mg abundances varied to improve the quality of the fit. The
results from the spectral fit are presented in
Table~\ref{tab:ps} and Figure~\ref{fig:ps}. We note that as for other
single WN stars the intrinsic (unabsorbed) X-ray emission from WR 7 is
dominated by the cool plasma component (kT~$= 0.6$ keV). It is
interesting to note that using the adopted stellar wind parameters and
distance to WR 7 (\S~\ref{sec:ngc2359}), we derive for the X-ray and
the wind luminosity L$_X = 1.18\times10^{32}$~erg~s$^{-1}$~ and 
L$_{wind} = (1/2)\dot{M} V_{wind}^2 = 3.25\times10^{37}$~erg~s$^{-1}$.
These numbers place WR 7 in accord with the correlation L$_X$ vs.
L$_{wind}$ established for presumably single WN stars (see Fig.10 in
\citealt{sk_10} and Fig.5 in \citealt{sk_12}).
Also, as for \wbb (\S~\ref{subsec:ngc2359}), the derived value for 
X-ray absorption is consistent with the optical extinction to WR 7.
We found no short-term variability in the \XMM data of WR 7 that 
is on timescales less than the effective exposures of $\sim 64$~ksec 
(pn) and $\sim 90$~ksec (MOS1,2; see \S \ref{sec:data}).

{\it [S3XJ067536].}
The \XMM spectra of this source show no features at energies of the
usually strong X-ray lines. We therefore adopted an absorbed power-law
model for the spectral fit. The corresponding results are given in
Table~\ref{tab:ps} and Figure~\ref{fig:ps}.
No short-term variability of [S3XJ067536] was detected on
timescales less than the effective exposures of the pn and MOS data.

\section{Discussion}
\label{sec:discussion}
One of the basic results from the analysis of the \XMM data on \wbb is
that the X-ray emission of this WBB is rather soft and originates from 
a plasma with a relatively low temperature (see Fig.~\ref{fig:wbbspec} 
and Table~\ref{tab:ngc2359}). As the stellar wind velocity is very
high (see \S~\ref{sec:ngc2359}), the expected shock velocity is high
too, thus the postshock plasma temperature should be at least an
order of magnitude larger than the value deduced from the analysis of
the X-ray spectrum of \wbbe. In general, this is an indication of
efficient electron thermal conduction operating in this object that 
could cause `evaporation' of the gas from the outer (optical) nebula 
into the hot bubble, thus, lowering the gas temperature in its interior 
(e.g., \citealt{weaver_77}).

Unfortunately, due to the low statistics of the spectrum,
we were not able to fit for the abundances which could 
have helped us draw some conclusion about the origin of the hot gas in 
\wbbe. 
We recall that such an analysis was possible for another WBB 
(NGC~6888) whose spectrum had a very good photon statistics. It
provided the first observational evidence that the hot gas in NGC~6888
originates from the outer cold optical shell \citep{zhp_11}.
Therefore we adopted values typical for the optical nebula in the fits
to the X-ray emission from \wbb (see \S~\ref{subsec:ngc2359}).
However, a model fit to the spectrum of \wbb was successful 
even if typical WN abundances \citep{vdh_86} were adopted for the 
hot gas in \wbbe. We note that this fit is also statistically 
acceptable, $\chi^2$/dof~$= 48/47$, and the plasma temperature,
kT~$= 0.25^{+0.02}_{-0.02}$, is similar to that from
the fit with abundances typical for the 
optical nebula (see Table~\ref{tab:ngc2359}).
All this confirms that the limited photon statistics in the
X-ray spectrum of \wbb does not allow for deriving some valuable
information about the elemental abundances in its hot interior.
Nevertheless, the low plasma temperature in \wbbe, a likely indication
of efficient electron thermal conduction, motivates attempts to test
the results from hydrodynamic modelling of WBB with the observations of
\wbbe.

Before doing this, a few basic details of the physics of WBBs 
around massive
stars are worth recalling. A WBB forms when the stellar wind of a
massive star interacts with the circumstellar matter. In the case of
WR stars, the fast massive wind collides with the slow wind emitted in
the previous stage of the WR evolution, i.e., a RSG (red
supergiant) or a LBV (luminous blue variable) wind. An important
feature of such an interaction is that both gas flows have similar
density profile ($\rho_{wind} \propto 1/r^2$) which results in a 
structure (WBB) expanding with constant velocity. Thus, the dynamical 
age of a WBB around a WR star is practically equal to its physical 
age. The basic physical quantities that determine the global
characteristics of such a WBB are the mass-loss rates and velocities
of the fast (WR) and the slow (RSG or LBV) winds. The fast wind 
parameters can be derived from analysis of the emission of the central 
star of a WR WBB  in different spectral domains (e.g., UV, optical, 
radio). On the other hand, the slow wind characteristics are free 
input parameters for the hydrodynamic modelling, and observational
properties to match in the `startup' modelling stage are the size and
expansion velocity of the WBB.

To derive the global characteristics of \wbb and the physical 
parameters of its hot bubble, we used
the hydrodynamic code of A.V. Myasnikov that was developed for
modelling the physics of a standard WBB (for details see
\citealt{zhm_98}). This code allows modelling not only of adiabatic and 
radiative WBB but also of the case with efficient thermal conduction.
It correctly handles interacting gas flows with different chemical 
compositions.

For \wbb (see \S~\ref{sec:ngc2359}), the typical range of expansion 
velocity is 10-30\kms and its angular size corresponds to a linear 
radius of $2.40$~pc for the adopted distance of $3.67$~kpc to this 
object. The mass-loss rate and velocity of the fast stellar wind 
(the wind of the central star WR 7) are correspondingly 
$\dot{M}_{fw} = 4\times10^{-5}$\dotM and V$_{fw} = 1600$\kms.
As to the slow wind parameters ($\dot{M}_{sw}$, V$_{sw}$), we varied 
them in order to match the expansion velocity and the size of the 
nebula. 
We only recall that as the nebula is sweeping out through the material
of the slow wind, the expansion velocity of the nebula sets an
upper limit for the slow wind velocity.
Thus, we chose to run two sets of 
models with V$_{sw} =$~ 5 and 10\kms, respectively.

\begin{figure}
 \centering\includegraphics[width=3.5in,height=2.5in]{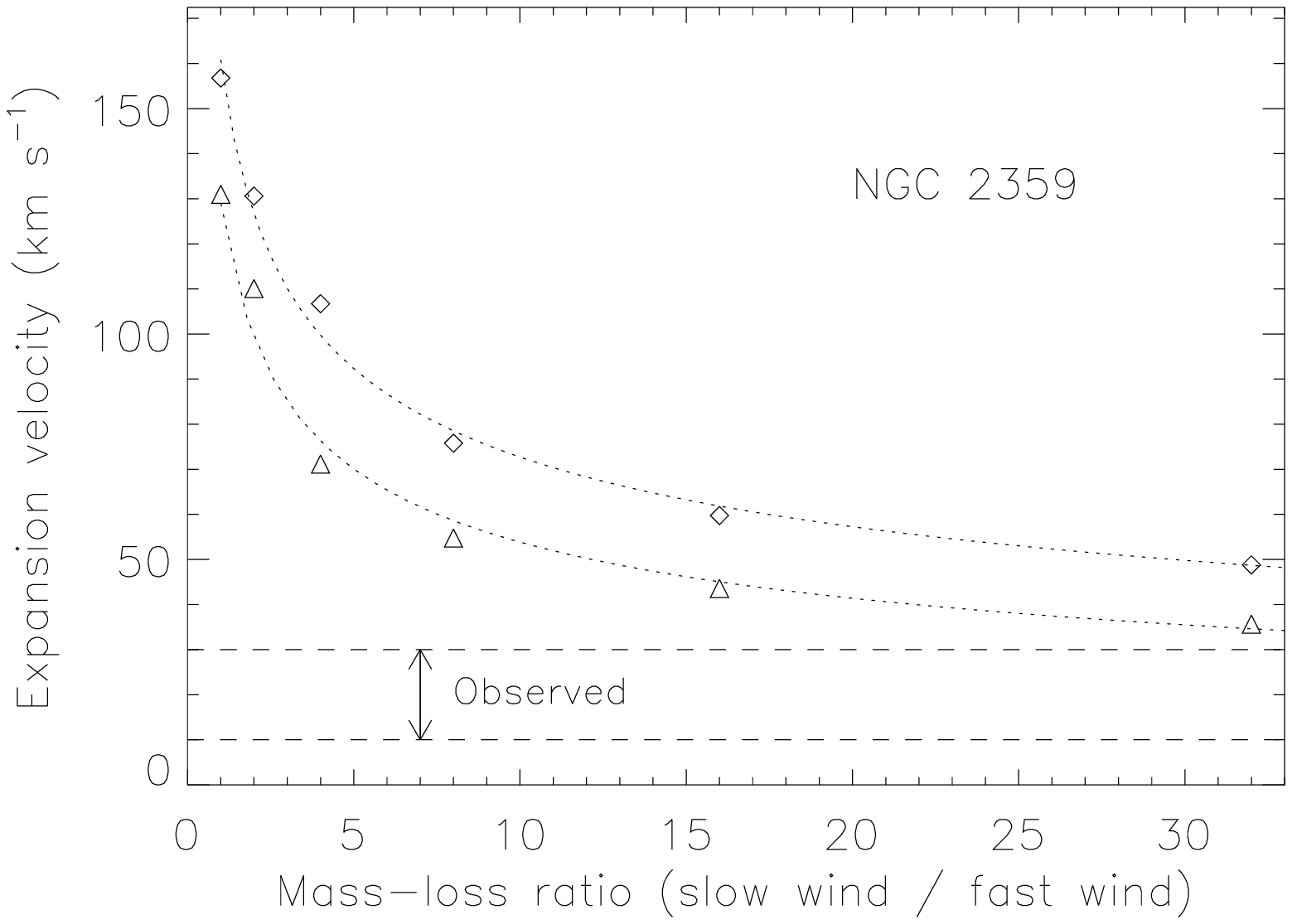}
\caption{Dependence of the expansion velocity of the nebula  on the
ratio of the mass loss of the slow wind to that of the fast wind in 
\wbbe. Results from two sets of hydrodynamic simulations are shown 
for a slow wind velocity of V$_{sw} =$~ 5 (triangles) and 10\kms 
(diamonds), respectively. For both sets, the fast wind parameters are 
$\dot{M}_{fw} = 4\times10^{-5}$\dotM and V$_{fw} = 1600$\kms. The 
two dashed lines mark the range of
values deduced from observations: $10 - 30$\kms \citep{chu_88}.
}
\label{fig:vexp}
\end{figure}

Figure~\ref{fig:vexp} presents the results for the expansion velocity
of the nebula as derived from the numerical simulations for various
values of the mass-loss ratio ($\dot{M}_{sw} / \dot{M}_{fw}$) of the 
stellar winds. We see that values close to the observed ones are
possible {\it only} for high mass-loss ratios: e.g., 
$\dot{M}_{sw} / \dot{M}_{fw} \geq 25 - 30$. For the nominal value of
the mass-loss rate of the fast wind of 
$\dot{M}_{fw} = 4\times10^{-5}$\dotM, this implies that the mass loss
of the slow wind should be very high:
$\dot{M}_{sw} \geq 10^{-3}$\dotM. We have to keep in mind that the
slow-wind phase in the evolution of the central star in a 
WBB should have lasted at least for a time period of 
$t_{sw} = t_{wbb}\times(V_{exp}/V_{sw} - 1)$\footnote{This relation
comes from a simple consideration that when the shock structure
expands to some distance at its age $t_{wbb}$, it should have found 
slow wind gas at the same distance from the star.}, 
where $t_{wbb}$ is the
age of the WBB and $V_{exp}$ is its expansion velocity. Given the
dynamical age (representative for the age of the nebula) of $\sim
100, 000$~years and the low expansion velocity  as deduced from 
observations (see \S~\ref{sec:ngc2359}), the standard physical 
picture of WBB suggests an unusually long slow-wind phase in the 
evolution of the central massive star in \wbb with extremely high 
mass-loss rate. 
This also implies a very large total mass loss of 
{\it $> 100$~solar masses} (!!). 
We note that such an evolutionary phase does not seem
to fit the standard evolutionary scenario for massive stars (e.g., see
\S 2.3.1 in \citealt{crowther_07}).

To further check the physical picture of WBB in the case of \wbbe, 
we made use of the WBB model with efficient thermal conduction.
We calculated the X-ray emission from the hot bubble for the case
with nominal values for the fast-wind parameters and the mass-loss
rate and velocity of $\dot{M}_{sw} = 1.28\times10^{-3}$\dotM 
($\dot{M}_{sw} / \dot{M}_{fw} = 32$) and $V_{sw} = 10$\kms. It was
done in the same way as in \citet{zhp_11} for the WBB NGC~6888.
Using the results from the one-dimensional hydrodynamic model, 
we calculated the 
distribution of emission measure of the hot bubble and the 
corresponding normalization parameter for the model spectrum in XSPEC 
($ norm = 10^{-14} \int n_e n_H dV / 4 \pi d^2$). 
Then, the actual fit to the X-ray spectrum showed if the 
theoretical emission measure gave the correct value for the observed 
flux from \wbbe. The result was that the theoretical X-ray emission
matched the shape of the observed spectrum acceptably well but it
overestimated the observed flux by more than {\it two orders of
magnitude}!
 
\begin{figure}
\centering\includegraphics[width=2.25in,height=3.01338in,angle=-90]
{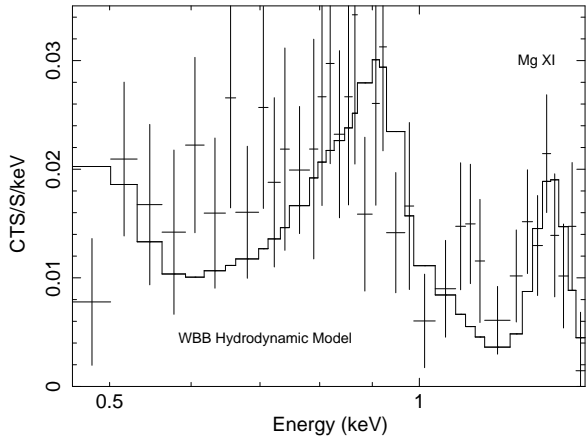}
\caption{
The \wbb background-subtracted spectrum as in Figure~\ref{fig:wbbspec}
overlaid with the best-fit model from the one-dimensional hydrodynamic 
simulations of conductive WBB with mass loss reduced by a factor of 
$\sim 13$ (see text for details).
}
\label{fig:hydrospec}
\end{figure}

To reconcile this flux discrepancy, it was necessary to reduce the
amount of X-ray emitting plasma in the hot bubble. 
To do so, we consecutively reduced the mass-loss rates of the fast and 
slow winds in a series of numerical simulations.
The ratio of mass-loss rates and the wind velocities were kept 
unchanged which resulted in no change of the global geometry of the 
shock structure.
The case with reduced mass-loss rate for the fast and slow winds of
$\dot{M}_{fw} = 3.0\times10^{-6}$\dotM and
$\dot{M}_{sw} = 9.6\times10^{-5}$\dotM provided the correct observed
flux. The corresponding fit ($\chi^2$/ dof~$= 47/47$) to the
\XMM spectrum of \wbb is shown in Figure~\ref{fig:hydrospec}.
We note that the derived X-ray absorption,
N$_{H} = 5.06^{+0.54}_{-0.48}\times 10^{21}$ cm$^{-2}$,
is consistent with the one from the spectral fit with the
discrete-temperature optically-thin plasma model 
(see Table~\ref{tab:ngc2359}).
Finally, we mention that a better match to the observed spectrum is
possible in the (0.55 - 0.65 keV) energy range if the oxygen abundance
is allowed to vary. However, as discussed above, the quality of the
data does not allow for getting some decent constraints on the
elemental abundances. Moreover, the varied oxygen abundance does not
alleviate the problem with the flux discrepancy, that is with the
requirement for a highly reduced mass-loss rate of the fast wind.

We thus see that as in the case of NGC~6888 \citep{zhp_11} the standard
physical model of a WBB can in general explain the properties of the
X-ray emission from \wbbe, provided the mass-loss rate of its central
star (WR 7) is considerably smaller (by a factor of $\sim 10$) than
the currently accepted value.
Future X-ray observations with higher sensitivity and lower detector
background will be very important for constraining the X-ray 
properties of \wbbe. This, in turn, will help us build a more 
realistic physical picture of this wind-blown bubble. Along these
lines, we mention that the low expansion velocity of \wbb poses
a problem for the standard physical model of WBBs, and this calls for
adopting a global approach to the study of this object. Namely, new
and more accurate measurements of the physical characteristics of the 
optical nebula, the central star itself, and the X-ray properties of 
\wbb should find their place in a `unified' picture which is capable
of explaining all of them in a self-consistent way.

\section{Conclusions}
\label{sec:conclusions}

In this work, we presented the \XMM data of \wbb which provide the 
first detection of this object in X-rays. We analysed the X-ray
properties of this wind-blown bubble and of its central star. 
The basic results and conclusions are as follows.

(i) As in the case of the other two WBBs detected in X-rays (S308 and 
NGC 6888), the X-ray spectrum from \wbb is rather soft and most of the
emission is in the 0.3 - 1.5 keV energy range. No appreciable emission
was detected at energies above 2 keV. The spectrum is well represented
by the emission from a relatively cool plasma with a temperature of 
kT~$\sim 0.2$~keV.

(ii) The central star of the nebula, WR 7, is detected in X-rays.
The spectrum of this WN4 object is well represented by the emission 
from a two-temperature plasma with a cool component of 
kT~$\sim 0.6$~keV and a hot component of kT~$\sim 2.7$~keV. 
The X-ray properties of WR 7 are thus similar to those of
other presumably single WN stars detected in X-rays.

(iii) The low plasma temperature in \wbb is a sign of efficient
electron thermal conduction operating in this object.
A direct comparison (in XSPEC) between the one-dimensional 
hydrodynamic model of conductive WBB and the X-ray spectrum of \wbb 
suggests a reduced mass-loss rate ($\sim 3\times10^{-6}$\dotM) of 
the central star in order to provide the correct value of the 
observed flux. 
We note that this figure is atypically low for the stellar wind in 
Wolf-Rayet stars. We thus think that the best way to build a
self-consistent physical picture of \wbb is to carry out
a global modelling of the entire system: central star, optical
nebula and the hot bubble.

\section{Acknowledgments}
This research has made use of the NASA's Astrophysics Data System, and
the SIMBAD astronomical data base, operated by CDS at Strasbourg,
France.
The optical image in Figure~\ref{fig:images} is downloaded from
SIMBAD.
The author is grateful to an anonymous referee for the valuable 
comments and suggestions.

\bsp

\label{lastpage}

\end{document}